\documentclass[preprint,aps,prc,showpacs,nofootinbib]{revtex4}
\usepackage{epsfig}
\usepackage{graphicx,amsmath}

\newcommand{\dd}{\mbox{d}}

\newcommand\ba{\begin{eqnarray}}
\newcommand\ea{\end{eqnarray}}
\newcommand\be{\begin{equation}}
\newcommand\ee{\end{equation}}
\newcommand\nn{\nonumber}
\newcommand{\thg}{\theta_{\gamma}}

\begin{document}

\title{Comments on ISR method in modern experiment and influence of final state radiation}

\affiliation{\it JINR-BLTP, 141980 Dubna, Moscow region, Russian Federation}
\author{E. Kuraev}
\affiliation{\it JINR-BLTP, 141980 Dubna, Moscow region, Russian Federation}
\author{V.~V.~Bytev}
\affiliation{\it JINR-BLTP, 141980 Dubna, Moscow region, Russian Federation}
\author{E.~Tomasi-Gustafsson}
\affiliation{CEA,IRFU,SPhN, Saclay, 91191 Gif-sur-Yvette Cedex, France, and \\
Universit\'e Paris-Sud, CNRS/IN2P3, Institut de Physique Nucl\'eaire, UMR 8608, F-91405 Orsay, France }
\author{S.~Pacetti}
\affiliation{Dipartimento di Fisica, Universita` di Perugia,
         I.N.F.N. Sezione di Perugia,
      Via Pascoli, 06123 Perugia, Italy}

\date{\today}

\begin{abstract}
We study the effect of final state radiation in the process  $e^++e^- \to \bar p+ p$, in the kinematical conditions
of BaBar and BESIII experiment. We show that this effect could be large, in particular in the low $x$ region
($x$ is the photon energy fraction) and should be taken into account.

\end{abstract}
\maketitle

The reaction $e^+ + e^- \rightarrow N+ \bar N$ has been studied since many decades, similarly to the crossed reactions $p+\bar p\leftrightarrow e^+ + e^-$, as they are considered the simplest reactions which contain information on the nucleon structure. If one describes the interaction through one photon exchange, a simple and elegant formalism allows to express the cross section and all polarization observables in terms of two form factors (FFs), which are real in the scattering channel and complex in the annihilation channels, due to unitarity.

At fixed energy $e^+e^-$ colliders, the emission of a hard photon in the initial state, $e^+ + e^- \rightarrow N+ \bar N+\gamma$, allows the measurement of the non radiative processes, $e^+ + e^- \rightarrow N+ \bar N$ over a range of $N\bar N$ energy from the threshold,
$w=2m_N$ to the full $e^+e^-$ center of mass energy, $w=\sqrt{s}$.

In \cite{BaBar}, based on the work of \cite{BM}, the differential cross section for the radiative process, integrated over the nucleon momenta, was factorized in a function which depends on the photon kinematical variables multiplied by the annihilation cross section for $e^+ + e^- \rightarrow N+ \bar N$:
\be
\displaystyle\frac{d^2\sigma_{e^+e^-\to \bar pp\gamma}(w)}{dwd\cos\thg }=
\displaystyle\frac{2w}{s}W(s,x,\thg )\sigma_{p\bar p}(w).
\label{eq:eqisr}
\ee
The total cross section for the annihilation process as function of the $p\bar p$ system invariant mass $w$ is :
\be
\sigma_{p\bar p}(w)=\frac{4\pi\alpha^2\beta C}{3w^2}
\left[|G_M(w)|^2+
\frac{2M^2}{w^2}|G_E(w)|^2\right ].
\label{eq:eqspp}
\ee
$\beta=\sqrt{1-4M^2/w^2}$ is the proton velocity, $M$ is the proton mass, 
$C=y/(1-e^{-y})$, and $y=\pi\alpha M/(\beta w)$ is the Coulomb correction factor. Note that very close to threshold, when $\beta\to 0$,  $y$ should be multiplied by a factor of two \cite{Ho97}.
The function $W(s,x,\thg )$ depends on the kinematical variables of the hard photon only. The expression used in \cite{BaBar} is: 
\be
W(s,x,\thg)=\displaystyle\frac{\alpha}{\pi x}\left ( \displaystyle\frac{2-2x+x^2}{\sin^2\thg}-\displaystyle\frac{x^2}{2}\right ).
\label{eq:babar}
\ee
Note that at zero photon emission angle Eq. (\ref{eq:babar}) does not apply because it neglects terms in $(m^2/s)$ which become important at small angles, see Ref. \cite{Ben99,Baier}.

In Eq. (\ref{eq:eqisr}), the factorization of the photon variables allows to extract the time-like proton form factors. The radiative emission from the proton may spoil the factorization hypothesis. At our knowledge, this has never been taken into account.

The purpose of this work is to estimate the final state radiation (FSR) contribution to the  $e^++e^-\to\bar p +p $ process: 
\begin{gather}
e^-(p_1)+e^+(p_2)\to \bar p(p_3)+p(p_4)+\gamma(k),
\end{gather}
in the kinematics of present experiments (BaBar and BESIII), including the electromagnetic proton FFs. The particle momenta are indicated in parenthesis.

\begin{figure}
\includegraphics[scale=.6]{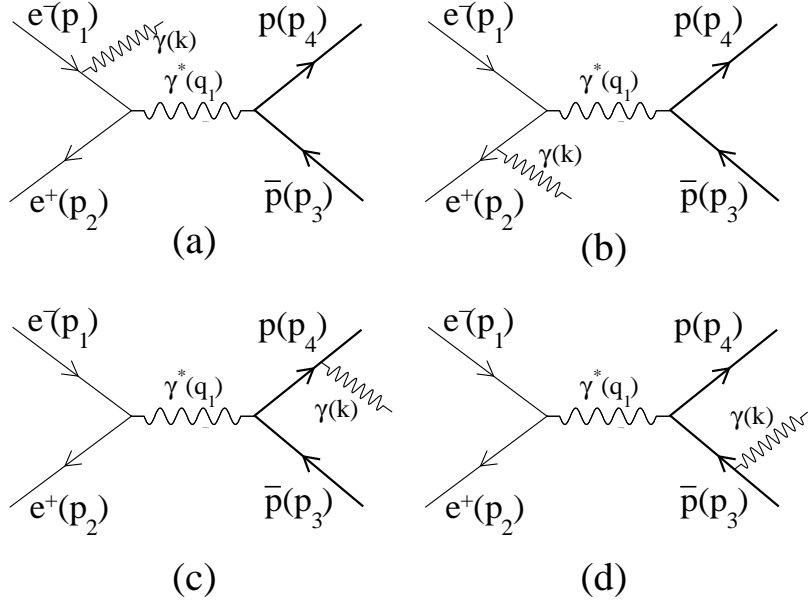}
\caption{ ISR and FSR diagrams for $e^++e^-\to \bar p+p+\gamma$, in Born approximation.}
\label{fig:Feyn}
\end{figure}

For initial state radiation (ISR), we derive expressions which hold in Born approximation. For FSR the present formalism  is correct for soft photon emission, and contains an
estimation of hard photon contribution. Virtual corrections are not included. The calculated diagrams are shown in Fig \ref{fig:Feyn}.

In the considered experiments where the detection is symmetrical, the contribution
of ISR-FSR interference vanishes, due to the integration over the final state of the proton pair (as it is antisymmetric, with respect to  $p_3\leftrightarrow p_4$ exchange).

The ISR cross-section, taking into account proton FFs, can be written in the form:
\be
\dd\sigma^{ISR}=\frac{4}{4 I}\sum_{pol}T_{\mu\nu}^{init,\gamma}\frac{\dd^3 k}{2\omega}\frac{(2\pi)^4}{(2\pi)^9}\frac{(4\pi\alpha)^3}{(q_1^2)^2}
\left (g^{\mu\nu}-\frac{q_1^\mu q_1^\nu}{q_1^2}\right )f(q_1^2)\frac{\dd^3 p_3}{2\varepsilon_3}\frac{\dd^3 p_4}{2\varepsilon_4}
\delta^4(p_1+p_2-p_3-p_4-k).
\ee
Here $q_1=p_3+p_4$ is the momentum of proton pair in the final  state, $I=\sqrt{(p_1p_2)^2-m_e^4}$ is the flux of initial particles,
 $T_{\mu\nu}^{init,\gamma}$ is the initial state radiation tensor:
\ba
\sum_{pol}T_{\mu\nu}^{init,\gamma}&=&-\frac{1}{4}Tr(\hat{p}_2-m_e)
\left (\gamma_\lambda\frac{-p_2+k+m_e}{-2\chi_2}\gamma_\mu+\gamma_\mu\frac{p_1-k+m_e}{-2\chi_1}\gamma_\lambda\right )
\nn\\
&&(\hat{p}_1+m_e)
\left (\gamma_\nu\frac{-p_2+k+m_e}{-2\chi_2}\gamma_\lambda+\gamma_\lambda\frac{p_1-k+m_e}{-2\chi_1}\gamma_\nu\right ).
\ea
Here and further we use:
\begin{gather}
\chi_i=kp_i, \,\,\, i=1\dots 4,\quad  q=p_1+p_2,\quad  q^2=s.
\end{gather}
Rearranging the phase volume of the proton pair and integrating the tensor of proton pair production (including FFs) in the reference frame where $\vec{q}_1=0$, we obtain:
\ba
\dd\Gamma_p&=&\frac{\dd^3 p_3}{2\varepsilon_3}\frac{\dd^3 p_4}{2\varepsilon_4}\delta^4(p_1+p_2-p_3-p_4-k)
=\frac{\pi \dd c_4}{4}\sqrt{1-\frac{4m_p^2}{q_1^2}},\quad c_4=cos(\widehat{k p_4} )
\nn\\
F_\mu&=&\gamma_\mu F_1(q_1^2)+\frac{F_2(q_1^2)}{2 m_p}\sigma^{\mu\lambda}q_1^\lambda,\quad
\tilde F_\mu=\gamma_\mu F_1^\star(q_1^2)-\frac{F_2^\star(q_1^2)}{2 m_p}\sigma^{\mu\lambda}q_1^\lambda,
\nn\\
f(q_1^2)\dd\Gamma_p&=&\frac{1}{4\cdot 3}
Tr(p_3+m_p)F_\mu(p_4-m_p)\tilde F_{\mu}\dd \Gamma_p,
\nn\\
f(q_1^2)\dd\Gamma_p&=&\frac{\pi q_1^2}{4\cdot 3}\sqrt{1-\frac{4 m_p^2}{q_1^2}}
\left [
-2|F_1(q_1^2)|^2\left (1+\frac{2m_p^2}{q_1^2}\right )+6Re[F_1(q_1^2)F_2^\star(q_1^2)]
\right .\nn\\
&& \left .-2|F_2(q_1^2)|^2\left (\frac{1}{8}\frac{q_1^2}{m_p^2}+1\right )\right ]\dd \Gamma_p.
\label{eq:5}
\ea
Collecting all factors we obtain formula for ISR including proton FFs:
\ba
\label{eq::7}
&\frac{\dd\sigma^{ISR}}{\dd c\dd \omega}&=\frac{\alpha^3\omega}{3s}\frac{1}{q_1^2}
\sqrt{1-\frac{4 m_p^2}{q_1^2}}
\\
&&
\biggl[
2|F_1(q_1^2)|^2\left(1+\frac{2m_p^2}{q_1^2}\right )-6Re [F_1(q_1^2)F_2^\star(q_1^2)]
+2|F_2(q_1^2)|^2\left(\frac{1}{8}\frac{q_1^2}{m_p^2}+1\right )
\biggr ]
\nn\\
&&
\biggl[
-m_e^2\left (\frac{1}{\chi_1^2}+\frac{1}{\chi_2^2}\right )(q_1^2+2m_e^2)+2\left(\frac{\chi_2}{\chi_1}+\frac{\chi_1}{\chi_2}\right )
+\left(\frac{1}{\chi_1}+\frac{1}{\chi_2}\right )(4m_e^2-2s)+\frac{s^2-4m_e^4}{\chi_1\chi_2}
\biggr],
\nn
\ea
where $\omega$ and $c=\cos(\widehat{p_1 k})$ are the photon energy and emission angle correspondingly.

A similar formula can be obtained for FSR. Here we neglect the proton FFs
connected with real photon emission, but include those connected with the virtually exchanged photon:
\begin{gather}
\dd\sigma^{FSR}=\frac{1}{4 I}\sum_{pol}T_{\mu\nu}^{init}\frac{\dd^3 k}{2\omega}\frac{(2\pi)^4}{(2\pi)^9}\frac{(4\pi\alpha)^3}{s^2}
\left(g^{\mu\nu}-\frac{q^\mu q^\nu}{q^2}\right)\frac{4}{3}f(s)\frac{\dd^3 p_3}{2\varepsilon_3}\frac{\dd^3 p_4}{2\varepsilon_4}
\delta^4(p_1+p_2-p_3-p_4-k).
\end{gather}
Here
\ba
\sum_{pol}T_{\mu\nu}^{init}&=&\frac{1}{4}Tr(\hat{p}_2-m_e)\gamma_\mu(\hat{p}_1+m_e)\gamma_\nu.
\nn\\
f(q_1^2)\dd\Gamma_p&=&\frac{1}{4\cdot 3}
Tr(p_3+m_p)\left(\gamma_\lambda\frac{p_3+k+m_p}{2\chi_3}F_\mu+F_\mu\frac{-p_4-k+m_p}{2\chi_4}\gamma_\lambda\right)
\nn\\
&&(p_4-m_p)\left(\tilde F_\mu\frac{p_3+k+m_p}{2\chi_3}\gamma_\lambda+\gamma_\lambda\frac{-p_4-k+m_p}{2\chi_4}\tilde F_\mu\right)
\dd\Gamma_p.
\ea
Integrating over the final proton state, see Eq. (\ref{eq:5}), we obtain for the 
FSR contribution:
\ba
\frac{\dd\sigma}{\dd \omega\dd c}=\frac{\alpha^3 \omega}{s^2 3}\sqrt{1-\frac{4 m_p^2}{q_1^2}}
\left [A\frac{16 q_1^2 s^2}{(s-q_1^2)^2 m_p^2}+B\frac{8s}{s-q_1^2 }\frac{1}{\beta}\ln\frac{1+\beta}{1-\beta}+2C\right],
\ea
where
\ba
A&=&-\frac{m_p^2}{s}|F_1(q_1^2)|^2\left (1+2\frac{m_p^2}{s}\right )+Re[F_1(q_1^2) F_2^\star]\frac{3 m_p^2}{s}
-|F_2(q_1^2)|^2\left (\frac{1}{8}+\frac{m_p^2}{s}\right )
\nn\\
B&=&|F_1(q_1^2)|^2\left (\frac{2}{x}-2+x-4\frac{m_p^2}{s}-8\frac{m_p^4}{x s^2}\right )
+Re[F_1(q_1^2) F_2^\star](q_1^2)\left(-\frac{6}{x}+6-x+12\frac{m_p^2}{sx}\right)
\nn\\
&&+|F_2(q_1^2)|^2\left(\frac{1}{4}\frac{s}{xm_p^2}-\frac{1}{4}\frac{s}{m_p^2}+\frac{1}{8}\frac{x s}{m_p^2}
+\frac{3}{2x}-2-\frac{4 m_p^2}{s x}\right)
\nn\\
C&=&-4|F_1(q_1^2)|^2-\frac{x s}{m_p^2}|F_2(q_1^2)|^2,~
\beta=\sqrt{1-\frac{4 m_p^2}{q_1^2}},\quad x=\frac{2\omega}{\sqrt{s}}.
\ea


For large values of $x$ or at small angle of photon emission, the final state radiation is strongly suppressed,
but for large-angle emission of photon and small fraction of photon energy, the contribution of
final state radiation is sizable.

\begin{figure}
\includegraphics[scale=1.0]{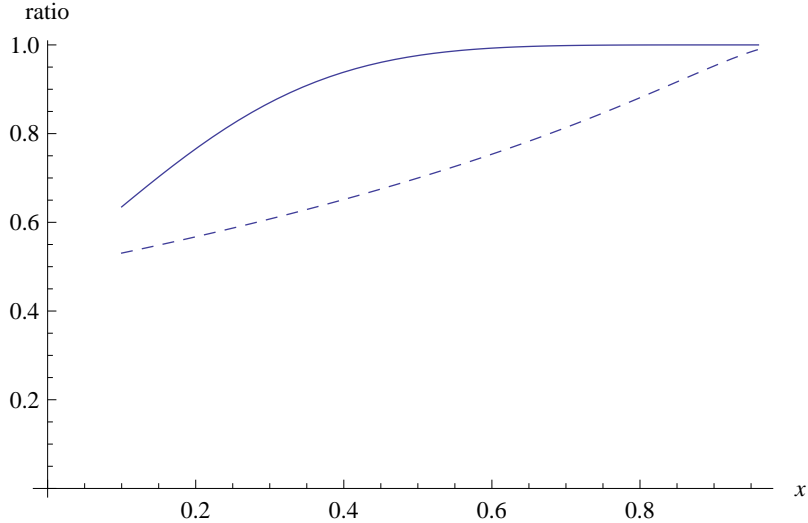}
\caption{Ratio of ISR cross-section to total cross section for point-like proton (dashed line) and including proton form factors (solid line) for BaBar kinematical conditions.}
\label{fig:fig1}
\end{figure}

It can be seen that for small $x$, the ratio $R$:
\ba
R=\frac{\sigma^{ISR}}{\sigma^{ISR}+\sigma^{FSR}}
\ea
deviates from unity for point-like proton, Fig. \ref{fig:fig1} (dashed line), in BaBar kinematical conditions. In Fig. \ref{fig:fig1} the same quantity is plotted including proton form-factors. the parametrization of FFs follows Ref.  \cite{TomasiGustafsson:2005kc} and it is based on an extension to the time-like region of the Iachello-Jackson-Lande model \cite{Ia03,Wa04}.
The photon emission angle is integrated in the region $(20,160)$ degrees. Let us remind that the BaBar detection corresponds to $s=111.6$ GeV$^2$, $E_{\gamma}>3$ GeV, $x>0.8$, and to the angular range (20,160) for the emitted proton.

It should be noted that here the integration over the proton and antiproton angles is done in the whole $4\pi$ region, but this does not change the ratio.

\begin{figure}
\includegraphics[scale=1.0]{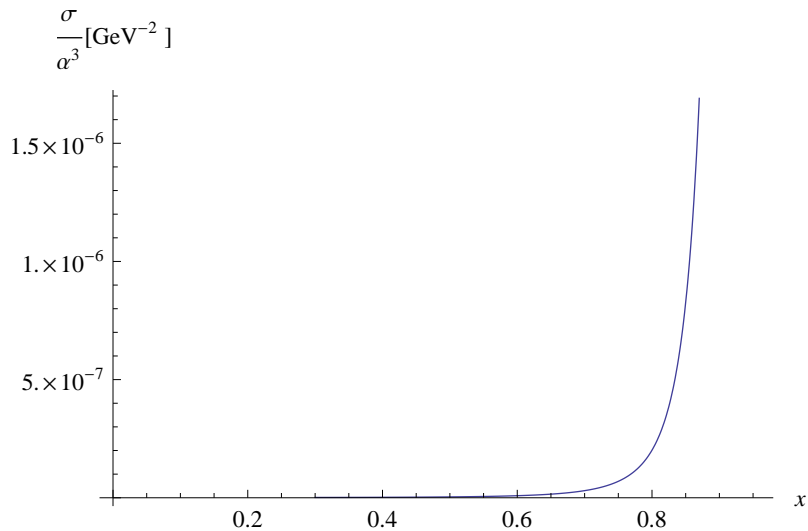}
\caption{ ISR cross-section for proton (BaBar conditions).}
\label{fig:fig2}
\end{figure}
One can see that for the BaBar experiment, almost all statistics appears in region $x>0.8$, (Fig. \ref{fig:fig2}), for which the correction
from final state is less that $1\%$ (see Fig \ref{fig:fig1}, solid line).

\begin{figure}
\includegraphics[scale=1.0]{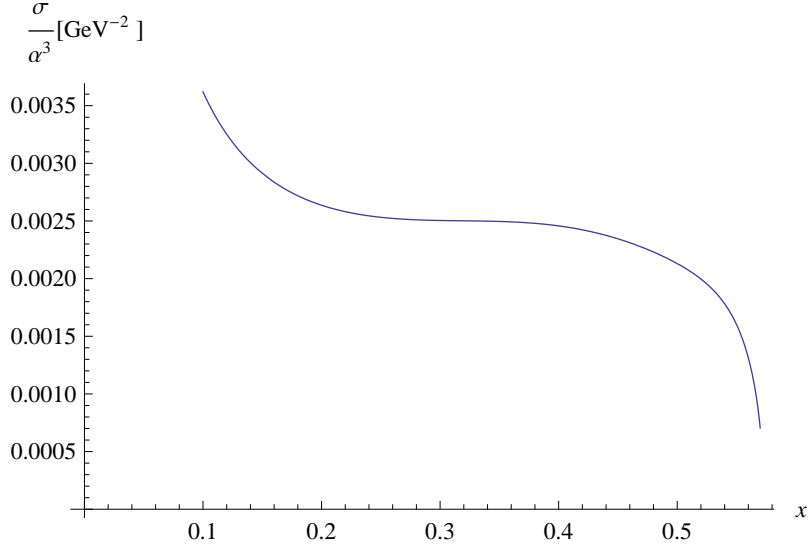}
\caption{ISR cross-section for BESIII conditions}
\label{fig:fig3}
\end{figure}

\begin{figure}
\includegraphics[scale=1.0]{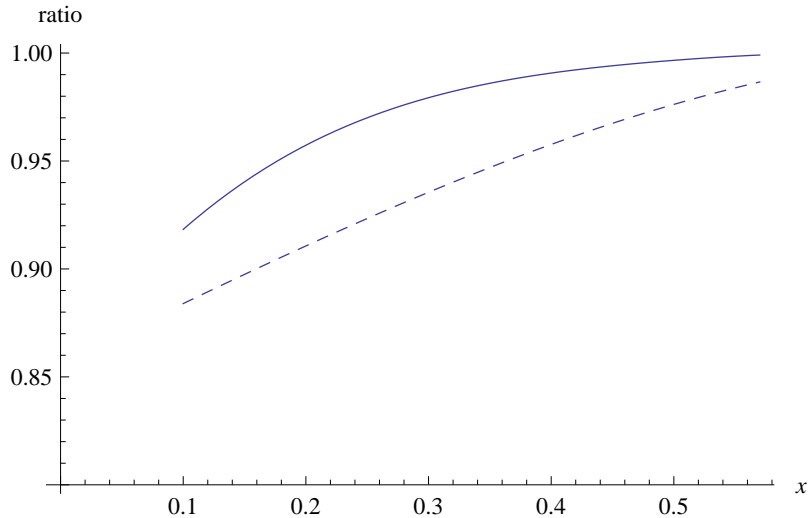}
\caption{Ratio of ISR cross-section to total cross section  for BESIII conditions, for point-like proton (dashed line) and including proton FFs  (solid line).}
\label{fig:fig4}
\end{figure}

For BESIII kinematical conditions ($\sqrt{s}=3$ GeV, minimal detection angle $5^\circ$), it can be seen that the cross section varies smoothly in the region of the detection (Fig. \ref{fig:fig3}),
and that the corrections from FSR could be sizable in the soft photon region (about $4\%$ at $x=0.2$) (Fig. \ref{fig:fig4}).

In conclusion, we have calculated initial and final state radiation for the process  $e^++e^-\to \bar p+p+\gamma$. We have given analytical formulas for the differential cross section on the proton variables, integrated over the angle of the hard photon. It appears that the factorization hypothesis holds for ISR but not for FSR. Therefore, for a reliable experimental result, one should either choose a region where FSR is negligible, or introduce corrections due to FSR to the cross section for the process  $e^++e^-\to \bar p+p+\gamma$, before factorizing out the terms due to ISR. Due to symmetry properties, it is possible to neglect the interference between ISR and FSR, when the detection of $\bar p  p$  is symmetrical.  

The integration over the photon angle in the experimental limits of BaBar, for point-like  protons, decreases drastically the contribution of ISR and makes the contribution of ISR and FSR comparable. Including the composite structure of the proton through FFs, decreases the relative contribution of FSR. This is due to an additional factor $1/(1-x)^4$, which increases the ISR contribution
(Fig. \ref{fig:fig1}).

In the experimental measurement of the cross-section for the reaction $e^++e^-\to \bar p+p$, using the ISR mechanism in BaBar conditions, 
the contribution of FSR for $x>0.8$ is negligible. For the BESIII  experiment FSR contribution could give sizable effect (about $4\%$). This conclusion does not depend on the chosen parametrization of the proton FFs, in the considered kinematical conditions. 
\section{Acknowledgments}
Prof. R. Baldini and Prof. M. Maggiora are acknowledged for stimulating discussions on form factor measurements at BES.
The french Groupement de Recherche Nucleon, is acknowledged for useful meetings and continuous support.

\end{document}